\documentclass[12pt]{article}
\font \greekb=cmmib10 scaled \magstep1
\newcommand{\sigmab}{\mbox{\greekb \char 27}}
\begin{document}

\begin{center}
{\large \bf Solar neutrino problem as evidence of new interaction}

\vspace{0.3 cm}

\begin{small}
\renewcommand{\thefootnote}{*}
L.M.Slad\footnote{slad@theory.sinp.msu.ru} \\
{\it Skobeltsyn Institute of Nuclear Physics,
Lomonosov Moscow State University, Moscow 119991, Russia}
\end{small}
\end{center}

\vspace{0.3 cm}

\begin{footnotesize}
A new concept is proposed to solve the solar neutrino problem, that is based on a hypothesis about the existence of a new interaction of electron neutrinos with nucleons mediated by massless pseudoscalar bosons. At every collision of a neutrino with nucleons of the Sun, its handedness changes from left to right and vice versa, and its energy decreases. The postulated hypothesis, having only one free parameter, provides a good agreement between the calculated and experimental characteristics of all five observed processes with solar neutrinos.

\end{footnotesize}

\vspace{0.3 cm}

\begin{small}

\begin{center}
{\bf 1. Introduction}
\end{center}

The present work is devoted to the substantiation of that the discrepancy between the predictions of the standard solar model (SSM) for the rates of a number of processes caused by solar neutrinos and the results of the appropriate experiments testify for the existence of a new, enough hidden interaction, which we name semi-weak. The postulated interaction involves the electron neutrino and the nucleons, but not the electron (at the tree level). Its carrier is a massless pseudoscalar boson. The product of the coupling constants of this boson to the electron neutrino and to the nucleons is smaller by three orders than the constants of electromagnetic and weak interactions, $\alpha$ and $g^{2}/4\pi$. At low energy values typical for solar and reactor neutrinos, the semi-weak interaction of a neutrino with nucleons refers to the cross sections of the order $10^{-36}$ cm$^{2}$, that are much larger than the standard cross sections of the $Z$-boson exchange processes. However, it is small enough to not-manifest itself in ordinary, nonunique situations. Its manifestation in the existing experiments with solar neutrinos is assisted with the circumstance that the Sun actually plays the role of part of the setup where every propagating neutrino undergoes about 10 collisions with nucleons. These collisions caused by the semi-weak interaction have the following three remarkable features.

First, at every collision with a nucleon, neutrino changes its handedness from left to right and vice versa. The ratio between the left- and right-handed electron neutrino fluxes at the Earth's surface coincides with the ratio between the probabilities of the even and odd number of collisions of a neutrino with nucleons before its exit from the Sun. Second, the total cross section of elastic scattering of the solar neutrino on a nucleon is practically independent on the neutrino energy. Therefore, we can consider that all solar neutrinos, irrespective of their energies, undergo on average the same number of collisions during their motion inside the Sun. Third, the relative fraction of the average energy loss of the neutrino in its collision with the nucleon of mass $M$ is proportional to its initial energy $\omega$: $\Delta \omega/\omega \simeq \omega/M$. Therefore, the energy of a neutrino from $p-p$-collisions remains almost unchanged, and the main factor influencing the results of experiments on the ${}^{71}{\rm Ga} \rightarrow {}^{71}{\rm Ge}$ transitions is the ratio of left- and right-handed electron neutrino fluxes at the Earth's surface. Since the cross sections of neutrino processes grow with increasing neutrino energy, the energy decrease of the tagged solar neutrinos due to their collisions with nucleons leads to the decrease of cross sections and of observed rates of processes involving solar neutrinos. The dependence of the cross sections on the neutrino energy is, undoubtedly, different for different processes, and, therefore, the difference in the effective fluxes of solar neutrinos from ${}^{8}{\rm B}$, corresponding to the rates of elastic scattering on electrons and of the deuteron disintegration with the production of the electron, is natural.

In the framework of the hypothesis about the existence of semi-weak interaction, the deuteron disintegration by neutral currents takes  a special place among the processes with solar neutrinos, as it is caused by two non-interfering sub-processes. One sub-process, with the participation of only the left-handed neutrinos, is a standard one and is due to the exchange of the $Z$-boson. Another, unordinary sub-process with the participation both left- and right-handed neutrinos is due to the exchange of the massless pseudoscalar boson. Due to the fact that, in the Lagrangian (\ref{1}), coupling constants of the postulated boson with the proton and the neutron are opposite, the cross section of the second sub-process is characterized by an additional factor $((M_{n}-M_{p})/M)^{2}$, so that it becomes comparable with the cross section of the first,  standard sub-process. It is quite natural that the effective solar neutrino flux corresponding to the total theoretical rate of the deuteron disintegration by the neutral currents, appears to be noticeably larger than the effective flux connected with the deuteron disintegration by the charged current.

The single parameter of the semi-weak interaction model describing the change of the energy spectrum of solar neutrinos during their motions from the production point to the outer borders of the Sun is taken to be the effective number of neutrino collisions with nucleons. Knowing this number, we estimate the value of the neutrino-nucleon coupling constant and use this estimate when calculating the cross section of the deuteron disintegration by the pseudoscalar neutral current of the solar neutrinos.

Our logical, analytical, and numerical analysis leads to the conclusion that the hypothesis about the existence of a new, semi-weak interaction is confirmed by a good agreement between the calculated and experimental characteristics of all observed processes with solar neutrinos: ${}^{37}{\rm Cl} \rightarrow {}^{37}{\rm Ar}$, ${}^{71}{\rm Ga} \rightarrow {}^{71}{\rm Ge}$, $\nu_{e} e^{-}\rightarrow \nu_{e} e^{-}$, $\nu_{e}D \rightarrow  e^{-}pp$, and $\nu_{e}D \rightarrow \nu_{e} np$.  The probability that this agreement reflects not the true nature of the things related to neutrinos, but a game of chance, seems to be negligible low. It is easy to make sure, that for last 40 years, no model or hypothesis has appeared in particle physics that, with only one free parameter, would give a good agreement with results of three or, furthermore, four or five, completely  different experiments.

Having a logically clear and elegant solution to the solar neutrino problem  presented in the current work, we should look more closely at the features of the interpretation of the solar neutrino experiments based on the neutrino oscillation hypothesis. Note only two of them. First, along with the hypothesis about solar neutrino oscillations in the vacuum based at least on two additional (massive) neutrinos and on two free parameters, the Wolfenstein--Mikheev--Smirnov mechanism is used. Second, in spite of hundreds of papers devoted to the solar neutrino oscillations, there is yet no publication in the world literature in which (1) a single-valued calculation procedure would be described for the probability that the solar neutrino remains to be electronic at the Earth surface, as function of its energy, (2) all free parameters of the procedure and their optimal values would be specified, and (3), that is most important, the summarized theoretical results for the rates of all five processes observed with solar neutrinos would be presented in comparison with the corresponding experimental results.

It is particularly noteworthy that all neutrino experiments in which the possibility of neutrino oscillation manifestations is investigated can be divided into three groups, respective to three practically independent branches of the oscillation model used in the interpretation of the data. Namely (see reviews \cite{1}, \cite{2}), experiments with solar neutrinos are mapped onto the parameters $\theta_{12}$ and $\Delta m_{21}$ (with large oscillation length and significant amplitude), experiments with antineutrinos from short-baseline reactors are mapped onto the parameters $\theta_{13}$ and $\Delta m_{31}$ (with short oscillation length and small amplitude), and experiments with atmospheric and accelerator neutrinos are mapped onto the parameter $\theta_{23}$ related to muon neutrinos. Consequently, the absence of manifestations of significant solar neutrino oscillations (generated by the branch with parameters $\theta_{12}$ and $\Delta m_{21}$) do not deny the possibility of small solar neutrino oscillations (generated by the branch with parameters $\theta_{13}$ and $\Delta m_{31}$) and the possibility of suitable oscillations in the neutrino experiments of other types, and vice versa.

A special place with respect to the solar neutrino experiments is occupied only experiments with reactor antineutrinos at KamLAND, because the interpretation of those and other experiments based on the oscillation model gives the values of the parameters $\theta_{12}$ and $\Delta m_{21}$ with the same order of magnitude.
The new interaction, which provides an elegant solution to the solar neutrino problem, has no attitude to emergence of the difference between expected and observed results in KamLAND. Recognizing the power of probabilistic regularities, the origin of these differences should be sought primarily in the omissions in the setting up and processing of the experiment. One such omission, noted in our work \cite{3} and based on indisputable numerical estimates, is due to the attenuation of fluorescent light in the liquid scintillator. Based on fragmentary experimental results of Tajima (subsequently a member of the KamLAND collaboration) \cite{3a}, we have found that the intensity of the resulting fluorescent signals at their propagation along the diameter of the spherical fiducial volume of the KamLAND liquid scintillator decreases by about five times. Since such a significant attenuation of the fluorescent light is not taken into account in theoretical calculations of observability of inverse beta-decay events, this leads to an uncontrolled loss of such events at KamLAND. 

\begin{center}
{\bf 2. Hypothesis about the existence of a massless pseudoscalar boson and its interaction}
\end{center}

In our solution of the solar neutrino problem, we are based on the
logically clear methods of the classical field theory.

We consider that the neutrino of each sort is described, similarly to the electron, by a bispinor representation of the proper Lorentz group, and its field obeys the Dirac equation. We note that all solutions with positive energy of the massless free Dirac equation, of which two (left-handed and right-handed) can be taken for basic ones, describe various states of the same neutrino. If there is external scalar or pseudoscalar field interacting with the neutrino, then the left and right spinors of neutrino wave vector will both have nonzero values.

The different kinds of hypothetical interactions involving neutrino were considered repeatedly. One of them, proposed by us \cite{4}--\cite{6}, was connected with a hypothetical massless axial photon. Now we can assert with sufficient confidence, not going into detail, that it is impossible to solve the solar neutrino problem by means of the axial photon as an interaction carrier. In parallel to this, an assumption about interaction of a hypothetical massless scalar with Majorana neutrino was expressed \cite{7}--\cite{10}. Because the potential energy $V(r)$ of the standard long-range forces has the behaviour $\sim r^{-1}$ for large enough distance $r$, only exceedingly faint interaction of such a scalar with others fermions was admitted, practically neither influencing the results of Eotvos type experiments nor  the value of the electron magnetic moment and, thereby, the solar neutrino spectrum.

So, we suppose, that there exists a massless pseudoscalar boson $\varphi_{ps}$, whose interaction with an electron neutrino, a proton and a neutron is described by the following Lagrangian
\begin{equation}
{\cal L} = ig_{\nu_{e}ps}\bar{\nu}_{e}\gamma^{5}\nu_{e}\varphi_{ps}+
ig_{Nps}\bar{p}\gamma^{5}p\varphi_{ps}-ig_{Nps}\bar{n}\gamma^{5}n\varphi_{ps},
\label{1}
\end{equation}
or by a similar Lagrangian with $u$- and $d$-quarks instead of proton $p$ and neutron $n$.

We intend neither to maintain, nor to deny the possibility to identify the boson $\varphi_{ps}$ with the Peccei--Quinn axion \cite{11}, \cite{12}, but we take into account the unsuccessfulness of experimental search for the axion and therefore postulate, for the sake of simplicity, the masslessness of the introduced boson, not excluding small enough values of its mass. The pseudoscalar boson $\varphi_{ps}$ is considered not interacting with the electron at tree level because it is impossible to fulfil simultaneously the three conditions: the values of coupling constants of this boson with the electron neutrino and the electron should be of the same order; the neutrino produced in the center of the Sun with the energy of the order 1 Mev should undergo at least one collision with some electron of the Sun; the contribution of such a boson to the value of the electron magnetic moment should not exceed the uncertainty limits admissible by the standard theory and experiments \cite{13}. We do not exclude that the interaction of boson $\varphi_{ps}$  with neutrinos of different sorts ($\nu_{e}$, $\nu_{\mu}$, and $\nu_{\tau}$) is nonuniversal, i.e., characterized by different coupling constants.

Note that the massless pseudoscalar field cannot manifest itself in Eotvos type  experiments, as the interaction of two nucleons mediated by it is similar to the magnetic interaction of spins, namely \cite{14}: $V(r) \sim r^{-3}[\sigmab_{1}\sigmab_{2}-3(\sigmab_{1}{\bf n})(\sigmab_{2}{\bf n})]$, where $\sigmab_{i}$
are the fermion spin matrices. Here is another simple reason in this respect.
For the standard long-range forces, the differential cross section of the elastic scattering of two charged particles described by the Rutherford formula has a pole at zero of the square of the momentum transfer, and the total cross section of such a scattering is infinite. In contrast to this, the differential cross section of the elastic scattering of two fermions caused by a massless pseudoscalar boson exchange has no pole, and total cross section of such a scattering is finite (see formulas (2) and (4)).

The interaction (\ref{1}), whose simple logical consequences are confirmed by all experiments with solar neutrinos, gives a bright realization of physics beyond the standard model, when some part of the general relativistically invariant Lagrangian has no traces of representations of any local symmetry group inherent to another part of the general Lagrangian. No relations existing in any gauge theory can be transferred to the Lagrangian (\ref{1}) and to the consequences from it at the tree level. And vice versa, the existence of the interaction (\ref{1}) has no effect on the construction of gauge models and on their consequences at the tree level.

\begin{center}
{\bf 3. Cross-sections and kinematics of the elastic neutrino-nucleon scattering}
\end{center}

The differential cross-section of the elastic scattering of the left- or right-handed electron neutrino with initial energy $\omega_{1}$ on a rest nucleon with mass $M$, obtained on the basis of Lagrangian (\ref{1}), is given by expression
\begin{equation}
d\sigma = \frac{(g_{\nu_{e}ps}g_{Nps})^{2}}{32 \pi M \omega_{1}^{2}}d\omega_{2},
\label{2}
\end{equation}
where $\omega_{2}$ is the scattered neutrino energy, which, as it results from the energy-momentum conservation law and from the formula (\ref{2}), can take evenly distributed values in interval
\begin{equation}
\frac{\omega_{1}}{1+2\omega_{1}/M} \leq \omega_{2} \leq \omega_{1}.
\label{3}
\end{equation}

The total cross-section of the elastic $\nu_{e}N$-scattering found from relations (\ref{2}) and (\ref{3}) is
\begin{equation}
\sigma = \frac{(g_{\nu_{e}ps}g_{Nps})^{2}}{16 \pi M^{2}}
\cdot \frac{1}{(1+2 \omega_{1}/M)}.
\label{4}
\end{equation}
It is practically independent on the energy $\omega_{1}$, since its maximal value is 18.8 MeV \cite{15}.

The first consequence of the interaction (\ref{1}) is that at each collision with a nucleon caused by an exchange of pseudoscalar boson, the neutrino changes its handedness from left to right and vice versa. Because of this, at the Earth's surface, one part of the solar neutrino flux has a left handedness, and the other part has a right handedness. The contribution from right-handed solar neutrinos to the charged current processes and to the elastic scattering on electrons is extremely small, since such neutrinos are not coupled with intermediate bosons of the standard model. They can be only coupled with very heavy intermediate bosons of the left-right symmetric model $W_{R}$ and $Z_{LR}$. The analysis of the nucleosynthesis in the early Universe \cite{16} and the electroweak fit \cite{17} give correspondingly the following estimate: $M_{W_{R}} > 3.3$ TeV and $M_{Z_{LR}} > 1.2$ TeV, if $g_{R} = g_{L}$. The last condition is automatically fulfilled in the initially $P$-invariant model of electroweak interactions \cite{18}. It is noteworthy that the both handednesses of solar neutrinos nevertheless manifest themselves, namely in the process of the deuteron disintegration by the neutral currents due to the exchange of massless pseudoscalar boson.

The second consequence of the interaction of the solar neutrino with a nucleon resulting from relations (\ref{2}) and (\ref{3}) is the neutrino energy decrease, on the average, by the value of
\begin{equation}
\Delta \omega_{1} = \frac{\omega_{1}^{2}}{M}\cdot \frac{1}{1+2\omega_{1}/M}.
\label{5}
\end{equation}
The formula (\ref{5}) shows that the relative single-shot change in the energy $\Delta \omega_{1}/\omega_{1}$ for solar neutrinos from ${}^{8}{\rm B}$ (their average energy equals 6.7 MeV) is by one order of magnitude higher than that for neutrinos from $p$ - $p$ (their maximum energy equals 0.423 MeV) and from ${}^{7}{\rm Be}$ with energy 0.384 and 0.862 MeV. Neutrinos from ${}^{8}{\rm B}$ play a main role in the ${}^{37}{\rm Cl} \rightarrow {}^{37}{\rm Ar}$ transitions and play a monopoly role in the elastic scattering on electrons and in the deuteron disintegration. At the same time, neutrinos from $p$-$p$ and from ${}^{7}{\rm Be}$ give a dominant contribution to the ${}^{71}{\rm Ga} \rightarrow {}^{71}{\rm Ge}$ transitions. Their energy changes very little after about ten neutrino-nucleon collisions, and the difference between the results of SSM and of the experiment is determined, almost completely, by the ratio between the fluxes of the left- and right-handed neutrinos. Since the experimental ${}^{71}{\rm Ga} \rightarrow {}^{71}{\rm Ge}$ transition rates are slightly smaller than half of the rate expected from SSM, we assume at the first approximation that the fluxes of left- and right-handed neutrinos at the Earth's surface are equal.

Due to collisions with nucleons, the solar neutrino from the moment of its production till the exit from the Sun makes Brownian motion in an inhomogeneous spherically symmetric medium. This motion is described, in principle, by some distribution $P_ {\beta}(n)$ in the number $n$ of collisions of a neutrino with the nucleons of the Sun, depending on the product of the coupling constants in  Lagrangian (\ref{1}), $\beta \equiv g_{\nu_{e}ps}g_{NPS}/4\pi$. Due to the relation (\ref{4}), we can consider that the distribution $P_{\beta}(n)$ is practically independent of the initial neutrino energy. Since solar neutrinos are produced in different areas, spaced from the Sun center at an average distance from 0.045 (from ${}^{8}{\rm B}$) till 0.14(from $hep$) of its radius \cite{15}, we should expect that the distribution $P_{\beta} (n)$ is wide enough, i.e. $|P_{\beta} (n+1)-P_{\beta} (n)| \ll 1$. From here, we obtain approximate equality of probabilities of even and odd number of neutrino-nucleon collisions inside the Sun and equality of fluxes of left- and right-handed neutrinos at the Earth's surface.

Being unable to accurately calculate the distribution $P_{\beta}(n)$, we replace, in the first approximation, the exact value of rate of the process $A$ with solar neutrinos, which includes the values of all partial rates $v_{A}(n)$ corresponding to the numbers of collisions of neutrinos with nucleons of the Sun $n$ with probabilities $P_{\beta}(n)$, by the rate $v_{A}(n_{0})$, corresponding to a single number of collisions of neutrinos $n_{0}$:
\begin{equation}
\sum_{n=0}^{+\infty} P_{\beta}(n) v_{A}(n) \rightarrow v_{A}(n_{0}).
\label{5a}
\end{equation}
The number $n_{0}$ from the relation (\ref{5a}) certainly depends on both the constant $\beta$ and the process $A$. In the considered approximation, we necessarily assume that the number $n_{0}$ is the same for all observed processes with solar neutrinos. A priori one cannot judge whether this assumption is good or bad. Fortunately, a comparison of theoretical and experimental results shows that such an assumption should be considered good.

An integer $n_{0}$, named the effective number of collisions, serves as the single free parameter of the model under consideration. It is not connected in any way with the ratio of neutrino handednesses at their exit from the Sun. The number $n_{0}$ describes the final energy distribution of a neutrino that had a fixed initial energy. As it was noted above, after one act of the elastic scattering of the neutrino on a rest nucleon, the initial fixed value of its energy is transformed into the evenly distributed energy interval (\ref{3}). After the second act of scattering, each energy value from this interval is transformed into its own interval of type (\ref{3}). Etc. 

Regarding the methods acceptable for calculations (say, in FORTRAN, as we did it), we have considered two variants to describe the energy distribution of neutrinos, having fixed initial energy $\omega_{i}$, after $n_{0}$ collisions with nucleons. In the first variant, the energy attributed to a neutrino after each  collision is equal to the mean value of the kinematic interval (\ref{3}), so that we have sequentially for zero, one, ..., $n_{0}$ collisions
\begin{equation}
\omega_{0,i}=\omega_{i}, \quad \omega_{1,i}=\omega_{0,i}\frac{1+\omega_{0,i}/M}
{1+2\omega_{0,i}/M}, 
\quad \ldots, \quad
\omega_{n_{0},i}=\omega_{n_{0}-1,i}\frac{1+\omega_{n_{0}-1,i}/M}
{1+2\omega_{n_{0}-1,i}/M}.
\label{6}
\end{equation}
In this variant, the final state of the neutrino is characterized by a single energy value given by the last term of the sequence (\ref{6}).

In the second variant, it is assumed that, as a result of each collision with a nucleon, the neutrino energy takes one of the two limiting values of the interval (\ref{3}) with equal probability and, by that, after $n_{0}$ collisions the initial level of energy $\omega_{i}$ turns into a set of $n_{0}+1$ binomially distributed values which elements are listed below:
\begin{equation}
E_{1,i} = \omega_{i}, \quad E_{2,i} = \frac{E_{1,i}}{1+2E_{1,i}/M}, \quad \ldots, \quad 
E_{n_{0}+1,i} = \frac{E_{n_{0},i}}{1+2E_{n_{0},i}/M}.
\label{7}
\end{equation}
Both variants yield close results. Thus, the replacement of the energy interval (\ref{3}) by three, four, etc. equiprobable values is inexpedient. We use everywhere only the second variant, which is more comprehensible in its logical plan than the first.

Let us turn now to specific experiments on registration of solar neutrinos.

\begin{center}
{\bf 4. The process $\nu_{e}+{}^{37}{\rm Cl} \rightarrow e^{-}+{}^{37}{\rm Ar}$}
\end{center}

The first experiment of this type \cite{19} has consisted in studying the process $\nu_{e}+{}^{37}{\rm Cl} \rightarrow e^{-}+{}^{37}{\rm Ar}$, having
the threshold energy 0.814 MeV. Now the experimental rate of such transitions is considered equal to $2.56 \pm 0.16 \pm 0.16$ SNU (1 SNU is $10^{-36}$ captures per target atom per second) \cite{20}. At the same time, theoretical calculations based on the standard solar model (SSM) give though different, but significantly greater values, for example: $7.9 \pm 2.6$ SNU \cite{15} and $8.5 \pm 1.8$ SNU \cite{24}.

We take from Refs. \cite{15}, \cite{21}, and \cite{22} the tabulated values of a number of quantities which are necessary to us for calculating the rates of transitions ${}^{37}{\rm Cl}\rightarrow {}^{37}{\rm Ar}$ and ${}^{71}{\rm Ga} \rightarrow {}^{71}{\rm Ge}$ induced by left-handed solar neutrinos.

We use the dependence of the cross-section of the process of neutrino absorption by chlorine on the neutrino energy $\omega$, presented in the table IX  and partly in the table VII of Ref. \cite{15}, and we assign for this cross-section a linear interpolation in each energy interval. We note the strong enough dependence of the mentioned cross-section on energy $\omega$ (expressed below in MeV) \cite{23}: $\sigma^{\rm Cl}(\omega)\sim \omega^{2.85}$, if 
$\omega \in [1, 5]$, and $\sigma^{\rm Cl}(\omega) \sim \omega^{3.7}$, if 
$\omega \in [8, 15]$. Therefore it is necessary to expect, that the decrease in energy of solar neutrinos as a result of their collisions with nucleons affects the rate of ${}^{37}{\rm Cl} \rightarrow {}^{37}{\rm Ar}$ transitions strongly enough.

The energy values of the neutrino from ${}^{8}{\rm B}$, spreading from 0 to about 16 MeV, are given in the table of Ref. \cite{21} in the form of set $\omega_{i}^{B}=i\Delta^{B}$, where $i = 1, \ldots, 160$, $\Delta^{B}=0.1$ MeV, and their distribution is expressed through probability $p(\omega_{i}^{B})$ of that neutrinos possess energy in an interval $(\omega_{i}^{B}-\Delta^{B}/2, \ \omega_{i}^{B}+\Delta^{B}/2)$. Each of the energy distributions in the interval $[0, \ 1.73]$ MeV for neutrinos from ${}^{15}{\rm O}$ and in the interval $[0, \ 1.20]$ MeV for neutrinos from ${}^{13}{\rm N}$ is presented in tables of Ref. \cite{22} for 84 points, and the distribution for neutrinos from $hep$ is given in the table of Ref. \cite{15} for 42 values of energy in the interval $[0, \ 18.8]$ MeV. The energy spectrum of neutrinos from ${}^{7}{\rm Be}$ has two lines 
$\omega_{1}^{Be}=0.862$ MeV ($89.7 \%$) and $\omega_{2}^{Be}=0.384$ MeV ($10.3 \%$), and from $pep$ has one line $\omega_{1}^{pep}=1.442$ MeV. For solar neutrino fluxes at the Earth surface, the values (in units of ${\rm cm}^{-2}{\rm s}^{-1}$) presented in Ref. \cite{24} are taken:
$\Phi({}^{8}{\rm B}) = 5.79 \times 10^{6}(1 \pm 0.23)$,
$\Phi({}^{7}{\rm Be}) = 4.86 \times 10^{9}(1 \pm 0.12)$,
$\Phi({}^{15}{\rm O}) = 5.03 \times 10^{8}(1^{+0.43}_{-0.39})$,
$\Phi(pep) = 1.40 \times 10^{8}(1 \pm 0.05)$,  
$\Phi({}^{13}{\rm N}) = 5.71 \times 10^{8}(1^{+0.37}_{-0.35})$,
$\Phi(hep) = 7.88 \times 10^{3}(1 \pm 0.16)$. For all that in the calculations, we use only the average values of the fluxes without involving uncertainty into any estimations or conclusions.

In view of the assumption that the fluxes of the left-handed neutrino at the Earth surface are equal to half of the above-mentioned fluxes, the formulas for calculating the contributions to the rate of transitions ${}^{37}{\rm Cl} \rightarrow {}^{37}{\rm Ar}$ caused by neutrinos from ${}^{8}{\rm B}$ and ${}^{7}{\rm Be}$ can correspondingly be presented in the form
\begin{equation}
V({}^{37}{\rm Cl} \ | \ {\rm B}) = 0.5 \Phi({}^{8}{\rm B})
\sum_{i=1}^{160}\Delta^{B}p(\omega_{i}^{B})
\sum_{n=1}^{n_{0}+1}\frac{n_{0}!}{2^{n_{0}}(n-1)!(n_{0}+1-n)!}
\sigma^{\rm Cl}(\omega_{n,i}^{B}), 
\label{8}
\end{equation}
\begin{equation}
V({}^{37}{\rm Cl} \ | \ {\rm Be}) = 0.5 \times 0.897 \Phi({}^{7}{\rm Be})
\sum_{n=1}^{n_{0}+1} \frac{n_{0}!}{2^{n_{0}}(n-1)!(n_{0}+1-n)!}
\sigma^{\rm Cl}(\omega_{n,1}^{Be}),
\label{9}
\end{equation}
where energy values $\omega_{n,i}^{B}$ and $\omega_{n,1}^{Be}$ are given by the formula (\ref{7}), in which the quantity $\omega_{i}$ needs to be set equal to $\omega_{i}^{B}$ and $\omega_{1}^{Be}$ accordingly. The contributions from neutrinos from ${}^{15}{\rm O}$, ${}^{13}{\rm N}$, and $hep$ are calculated by a formula similar to (\ref{8}), and the contribution from $pep$ does by a formula similar to (\ref{9}).

Before we proceed to calculations using formulas of the type (\ref{8}) and (\ref{9}) with nonzero value of the number of neutrino-nucleon collisions $n_{0}$, we check how much are the results, obtained with using the tabulated and interpolated values for the cross-section $\sigma^{\rm Cl}(\omega)$ under the condition of free motion of neutrinos in the Sun, close to what are obtained in Ref. \cite{15} on the basis of more precise calculations, though at slightly different spectra and flux values. This comparison is reflected in table 1.

The calculations concerning all of the discussed below processes give in their integrity the best agreement with experimental results if $n_{0} = 11$. To display the dependence of the theoretical results on the integer $n_{0}$, we present them in table 1 and in the subsequent for the two values of $n_{0}$, 10 and 11.

\begin{center}
\begin{tabular}{lccccccc}
\multicolumn{8}{c}{{\bf Table 1.} The rate of transitions ${}^{37}{\rm Cl} \rightarrow {}^{37}{\rm Ar}$ in SNU.} \\ 
\hline
\multicolumn{1}{l}{} 
&\multicolumn{1}{c}{${}^{8}{\rm B}$} 
&\multicolumn{1}{c}{${}^{7}{\rm Be}$}
&\multicolumn{1}{c}{${}^{15}{\rm O}$}
&\multicolumn{1}{c}{$pep$}
&\multicolumn{1}{c}{${}^{13}{\rm N}$}
&\multicolumn{1}{c}{$hep$}
&\multicolumn{1}{c}{Total} \\ 
\hline
SSM \cite{15} & 6.1 & 1.1 & 0.3 & 0.2 & 0.1 & 0.03 & 7.9 \\
Interpolations, & & & & & & & \\
no interactions & 6.21 & 1.05 & 0.35 & 0.22 & 0.09 & 0.02 & 7.94 \\
Interaction (\ref{1}), & & & & & & & \\
$n_{0} = 10$ & 2.05 & 0.44 & 0.17 & 0.11 & 0.04 & 0.01 & 2.82 \\
Interaction (\ref{1}), & & & & & & & \\
$n_{0} = 11$ & 1.97 & 0.43 & 0.17 & 0.11 & 0.04 & 0.01 & 2.72 \\
\hline
\end{tabular}
\end{center}

\begin{center}
{\bf 5. The process $\nu_{e}+{}^{71}{\rm Ga} \rightarrow e^{-}+{}^{71}{\rm Ge}$}
\end{center}

Let us turn now to the process $\nu_{e}+{}^{71}{\rm Ga} \rightarrow e^{-}+{}^{71}{\rm Ge}$ having the threshold energy 0.233 MeV. The latest experiments have given the following values for the rate of this process: 
$65.4^{+3.1}_{-3.0}{}^{+2.6}_{-2.8}$ SNU \cite{25} and $62.9^{+6.0}_{-5.9}$ SNU \cite{26}. From the theoretical results, which are worthy mentioning, we note two: $132^{+20}_{-17}$ SNU \cite{15} and $131^{+12}_{-10}$ SNU \cite{24}.

We use the neutrino energy dependence of the cross-section of the process with gallium $\sigma^{\rm Ga}(\omega)$ presented in the table II of Ref. \cite{22}, and  interpolate it inside the each interval by a linear function. In addition to the information written above on the neutrino fluxes, some data about neutrinos from $p$-$p$ is still required. The tabulated energy spectrum of such neutrinos, available in Ref. \cite{15}, spreads from 0 to 0.423 MeV and is given by a set of 84 points. The flux of solar neutrinos from $p$-$p$ at the Earth surface is taken equal to $\Phi(pp) = 5.94 \times 10^{10} (1 \pm 0.01)$ ${\rm cm}^{-2}{\rm s}^{-1}$ \cite{24}.

We calculate the contributions to the rate of transitions ${}^{71}{\rm Ga} \rightarrow {}^{71}{\rm Ge}$ brought by solar neutrinos from $p$-$p$, ${}^{8}{\rm B}$, ${}^{15}{\rm O}$, ${}^{13}{\rm N}$, and $hep$, with the formula similar to (\ref{8}), and the contributions brought by two lines of ${}^{7}{\rm Be}$ and by one line of $pep$, with the formula similar to (\ref{9}). The results of calculations are presented in table 2.

\begin{center}
\begin{tabular}{lcccccccc} 
\multicolumn{9}{c}{{\bf Table 2.} The rate of transitions ${}^{71}{\rm Ga} \rightarrow {}^{71}{\rm Ge}$ in SNU.} \\
\hline
\multicolumn{1}{l}{}
&\multicolumn{1}{c}{$p$-$p$}
&\multicolumn{1}{c}{${}^{7}{\rm Be}$} 
&\multicolumn{1}{c}{${}^{8}{\rm B}$} 
&\multicolumn{1}{c}{${}^{15}{\rm O}$}
&\multicolumn{1}{c}{${}^{13}{\rm N}$}
&\multicolumn{1}{c}{$pep$}
&\multicolumn{1}{c}{$hep$}
&\multicolumn{1}{c}{Total} \\ 
\hline
SSM \cite{15} & 70.8 & 34.3 & 14.0 & 6.1 & 3.8 & 3.0 & 0.06 & 132 \\
Interpolations, & & & & & & & & \\
no interactions & 69.8 & 34.9 & 14.0 & 5.7 & 3.4 & 2.9 & 0.05 & 130.7 \\
Interaction (\ref{1}), & & & & & & & &  \\
$n_{0} = 10$ & 34.7 & 17.2 & 5.0 & 2.8 & 1.7 & 1.4 & 0.02 & 62.8 \\
Interaction (\ref{1}), & & & & & & & &  \\
$n_{0} = 11$ & 34.6 & 17.2 & 4.9 & 2.8 & 1.7 & 1.4 & 0.02 & 62.6 \\
\hline
\end{tabular}
\end{center}

The fact that the theoretical value of the rate of transitions ${}^{71}{\rm Ga} \rightarrow {}^{71}{\rm Ge}$ at $n_{0}=10$ agrees with the above-mentioned experimental values is an important evidence, firstly, in favour of our assumption about the approximate equality of the fluxes of the left- and right-handed solar neutrinos at the Earth surface and, secondly, in favour of the consequence (\ref{5}), resulting from Lagrangian (\ref{1}), about decreasing the relative neutrino energy change at a single-shot collision with decreasing energy.

\begin{center}
{\bf 6. The process $\nu_{e}+e^{-} \rightarrow  \nu_{e}+e^{-}$}
\end{center}

Let us turn to consideration of the process of elastic scattering of solar neutrinos on electrons $\nu_{e} e^{-} \rightarrow  \nu_{e} e^{-}$, taking into account the conditions and the results of experiments at Super-Kamiokande 
\cite{27}--\cite{29} and at the Sudbury Neutrino Observatory (SNO) \cite{30}--\cite{33}.

The differential cross-section of elastic scattering of the left-handed neutrino with initial energy $\omega$ on a rest electron with mass $m$ is given by the formula (see, for example, Ref. \cite{34}, \cite{35})
\begin{equation}
\frac{d\sigma_{\nu e}}{dE} = \frac{2G_{F}^{2}m}{\pi}\left[ g_{L}^{2}+g_{R}^{2}\left( 1-\frac{E-m}{\omega}\right)^{2} - g_{L}g_{R} \frac{m(E-m)}{\omega^{2}} \right] \equiv f_{\nu e}(\omega, E),
\label{10}
\end{equation}
where $E$ is the energy of the recoil electron. For the scattering of the electron neutrino, we have in the Weinberg--Salam model of electroweak interactions 
\begin{equation}
g_{L}= \frac{1}{2}+\sin^{2}\theta_{W}, \quad g_{R}=\sin^{2}\theta_{W},
\label{11}
\end{equation}
where it is necessary to set $\sin^{2}\theta_{W} = 0.231$.

On the basis of the energy-momentum conservation law, we obtain that the recoil electron can get the energy $E$ if the energy $\omega$ of the incident neutrino satisfies the condition
\begin{equation}
\omega \geq \frac{E-m+\sqrt{E^{2}-m^{2}}}{2} \equiv h_{\nu e}(E).
\label{12}
\end{equation}

At setting up an experiment on the elastic neutrino-electron scattering, it is considered that a distinction between the true energy $E$ of the recoil electron and its reconstructed (effective) energy $E_{\rm eff}$ is given by the Gaussian probability density
\begin{equation}
P(E_{\rm eff}, E) = \frac{1}{\sigma \sqrt{2\pi}}\exp \left[ 
-\frac{(E_{\rm eff}-E)^{2}}{2\sigma^{2}}\right], 
\label{13}
\end{equation}
where the parameter $\sigma$, being a function of the energy $E$, depends on the features of an experimental set-up.

Since in all of the discussed experiments the lower limit $E_{c}$ for the reconstructed energy $E_{\rm eff}$ is introduced, and it is not less than 3 MeV, then the observable events are practically completely generated by the solar neutrinos from  ${}^{8}{\rm B}$, while the contribution from neutrinos from $hep$ is very small. The contribution to the rate of the scattering of neutrinos from ${}^{8}{\rm B}$ on electrons, when the reconstructed energy $E_{\rm eff}$ belongs to the interval from $E_{k}$ up to $E_{k+1}$ ($E_{k} \geq E_{c}$), is calculated according to the formula
$$V(\nu e \ | \ {\rm B} \ || \ [E_{k},E_{k+1}]) = 0.5 \Phi({}^{8}{\rm B})
\int_{E_{k}}^{E_{k+1}}dE_{\rm eff}
\int_{1 \ {\rm Mev}}^{16 \ {\rm Mev}}dE [P(E_{\rm eff},E)$$ 
\begin{equation}
\times \sum_{i=1}^{160} \Delta^{B}p(\omega_{i}^{B})\sum_{n=1}^{n_{0}+1} \frac{n_{0}!}{2^{n_{0}}(n-1)!(n_{0}+1-n)!}f_{\nu e}(\omega_{n,i}^{B},E) 
\theta(\omega_{n,i}^{B}-h_{\nu e}(E))],
\label{14}
\end{equation}
where $\theta(x)$ is the Heaviside step function. The contribution from neutrinos from $hep$ is found by the similar formula. The result of this or that experiment and, also, the theoretical calculation on the basis of the formula (\ref{14}) are expressed through the effective (either observable, or equivalent) flux of neutrinos from ${}^{8}{\rm B}$, $\Phi_{eff}({}^{8}{\rm B})$, which do not undergo any changes between the production place in the Sun and the experimental apparatus on the Earth. Connection between such a result or calculation and the effective flux is given by the following relation
$$V(\nu e \ | \ {\rm B}+hep \ || \ [E_{c}, 20 \ {\rm MeV}]) = 
\Phi_{eff}^{\nu e}({}^{8}{\rm B}) \int_{E_{c}}^{20 {\rm MeV}}dE_{\rm eff}
\int_{1 \ {\rm Mev}}^{16 \ {\rm Mev}}dE [P(E_{\rm eff},E)$$
\begin{equation}
\times \sum_{i=1}^{160} \Delta^{B}p(\omega_{i}^{B})
f_{\nu e}(\omega_{i}^{B},E) \theta(\omega_{i}^{B}-h_{\nu e}(E))].
\label{15}
\end{equation}

Let us notice here, that the Gaussian distribution (\ref{13}) has essential influence on the bin $[E_{k}, E_{k}+0.5 \ {\rm MeV}]$ distribution (\ref{14}) of the rate of $\nu e$-scattering events and has only small influence on the value of the effective neutrino flux found from equality (\ref{15}).

Some details of the experiments and of our calculations concerning the elastic scattering of solar neutrinos on rest electrons are presented in table 3, where  $T=E-m$.

\begin{center}
{{\bf Table 3.} Effective fluxes of neutrinos found from the process $\nu_{e} e^{-}\rightarrow \nu_{e} e^{-}$} \\
\begin{tabular}{lcccc}
\multicolumn{5}{c}{($E_{c}$, $E$, and $T$ are given in MeV, and the fluxes are in units of $10^{6}$ ${\rm cm}^{-2}{\rm s}^{-1}$).} \\ 
\hline
\multicolumn{1}{l}{References}
&\multicolumn{1}{c}{$E_{c}$} 
&\multicolumn{1}{c}{$\sigma$} 
&\multicolumn{1}{c}{Experi-}
&\multicolumn{1}{c}{Eq. (\ref{1}),} \\
&&& mental & $\Phi_{eff}^{\nu e}({}^{8}{\rm B})$, \\
& & & $\Phi_{eff}^{\nu e}({}^{8}{\rm B})$ & $n_{0}=10, 11$ \\
\hline
SK III \cite{29} & 5.0 & $-0.123+0.376\sqrt{E}+0.0349E$ &$2.32\pm 0.04\pm 0.05$ & $2.32 \hspace{0.4cm} 2.27$  \\
SK II \cite{28} & 7.0 & $0.0536+0.520\sqrt{E}+0.0458E$ &$2.38\pm 0.05{}^{+0.16}_{-0.15}$ & $2.07 \hspace{0.4cm} 2.00$  \\
SK I \cite{27} & 5.0 & $0.2468+0.1492\sqrt{E}+0.0690E$ &$2.35\pm 0.02\pm 0.08$ & $2.32 \hspace{0.4cm} 2.27$ \\
SNO III \cite{33} & 6.5 & $-0.2955+0.5031\sqrt{T}+ 0.0228T$ &$1.77^{+0.24}_{-0.21}{}^{+0.09}_{-0.10}$ & $2.08 \hspace{0.4cm} 2.01$ \\
SNO IIB \cite{32} & 6.0 & $-0.131+0.383\sqrt{T}+0.0373T$ &$2.35\pm 0.22\pm 0.15$ & $2.17 \hspace{0.4cm} 2.10$ \\
SNO IIA \cite{31} & 6.0 & $-0.145+0.392\sqrt{T}+0.0353T$ &$2.21^{+0.31}_{-0.26}\pm 0.10$ & $2.17 \hspace{0.4cm} 2.10$  \\
SNO I \cite{30} & 5.5 & $-0.0684+0.331\sqrt{T}+0.0425T$ &$2.39^{+0.24}_{-0.23}{}^{+0.12}_{-0.12}$ & $2.24  \hspace{0.4cm} 2.19$ \\
\hline
\end{tabular}
\end{center}

\begin{center}
{\bf 7. Deuteron disintegration by the charged current $\nu_{e}+D \rightarrow e^{-}+p+p$}
\end{center}

Let us dwell now on the process of deuteron disintegration by solar netrinos caused by the weak charged current interactions, $\nu_{e}D \rightarrow e^{-}pp$. The needed differential cross-section of this process, $d\sigma_{cc}/dE \equiv f_{cc}(\omega, E)$, as a function of energy $\omega$ of the incident left-handed electron neutrino and the energy $E$ of the produced electron is found in the tabulated form on a Web-site \cite{36}, which has resulted from the field-theoretical analysis of the $\nu D$-reaction presented in Ref. \cite{37}. In the tables of Ref. \cite{36}, the intervals in the neutrino energy are 0.2 MeV at $\omega \leq 10$ MeV and 0.5 MeV at $\omega > 10$ MeV, and the intervals in $E$ are varied in length. We resort to the linear extrapolation of the cross section in $\omega$ and $E$.

The kinematic condition for the neutrino energy needed for the deuteron disintegration has the form (see, for example, Ref. \cite{38})
\begin{equation}
\omega \geq E+B+\delta \equiv h_{cc}(E),
\label{16}
\end{equation}
where $B = 2.2246$ MeV is the binding energy of the deuteron, and $\delta = M_{p}-M_{n} = -1.2933$ MeV is the mass difference between the proton and the neutron.

We neglect the contribution to the deuteron disintegration from neutrinos from $hep$ and carry out the calculations of the rate of the process $\nu_{e}D \rightarrow e^{-}pp$ by the formula similar to (\ref{14}). The appropriate effective neutrino flux $\Phi_{eff}^{cc}({}^{8}{\rm B})$ is found from the relation similar to (\ref{15}).

The results of the SNO collaboration experiments and of our calculations are presented in table 4.
\begin{center}
{{\bf Table 4.} Effective fluxes of neutrinos found from \\ the process $\nu_{e}D \rightarrow  e^{-}pp$ ($E_{c}$ is given in MeV,} \\
\begin{tabular}{lccc}
\multicolumn{4}{c} {and the fluxes are in units of $10^{6}$ ${\rm cm}^{-2}{\rm s}^{-1}$).} \\ 
\hline
\multicolumn{1}{l}{References}
&\multicolumn{1}{c}{$E_{c}$}  
&\multicolumn{1}{c}{Experi-}
&\multicolumn{1}{c}{Eq. (\ref{1}),} \\
& & mental & $\Phi_{eff}^{cc}({}^{8}{\rm B})$, \\
& & $\Phi_{eff}^{cc}({}^{8}{\rm B})$ & $n_{0}=10, 11$ \\
\hline
SNO III \cite{33} & 6.5 & $1.67^{+0.05}_{-0.04}{}^{+0.07}_{-0.08}$ & $1.76 \hspace{0.4cm} 1.67$ \\
SNO IIB \cite{32} & 6.0 & $1.68^{+0.06}_{-0.06}{}^{+0.08}_{-0.09}$ & $1.87 \hspace{0.4cm} 1.78$  \\
SNO IIA \cite{31} & 6.0 & $1.59^{+0.08}_{-0.07}{}^{+0.06}_{-0.08}$ & $1.87 \hspace{0.4cm} 1.78$  \\
SNO I \cite{30} & 5.5 & $1.76^{+0.06}_{-0.05}{}^{+0.09}_{-0.09}$ & $1.96 \hspace{0.4cm} 1.88$  \\
\hline
\end{tabular}
\end{center}

Noting good agreement between the results of our calculations and the results of experiments concerning the effective neutrino fluxes which correspond to the events of elastic $\nu e$-scattering and the reaction $\nu_{e}D\rightarrow e^{-}pp$, we draw attention to the fact that the difference between theoretical values of the effective fluxes describing the two processes is due to the change in the shape of the solar neutrino spectrum because of their collisions with the nucleons of the Sun and due to different dependence of the cross sections of processes on the neutrino energy.

\begin{center}
{\bf 8. Deuteron disintegration by the neutral currents $\nu_{e}+D \rightarrow \nu_{e}+n+p$}
\end{center}

Experiments on the deuteron disintegration into a neutron and a proton caused by the neutral current of solar neutrinos are considered crucial for the theoretical interpretations. The consecutive results of such experiments in SNO are expressed with the following values of effective solar neutrino flux
$\Phi_{eff}^{nc}({}^{8}{\rm B})$ (in units of $10^{6}$ ${\rm cm}^{-2}{\rm s}^{-1}$): 
$$5.09^{+0.44}_{-0.43}{}^{+0.46}_{-0.43} \; \cite{30}, \quad 5.21 \pm 0.27 \pm 0.38 \; \cite{31}, \quad 4.94^{+0.21}_{-0.21}{}^{+0.38}_{-0.34} \; \cite{32}, \quad 5.54^{+0.33}_{-0.31}{}^{+0.36}_{-0.34} \; \cite{33}.$$ 
In view of our hypothesis about the existence of the interaction described by Lagrangian (\ref{1}), the deuteron disintegration into a neutron and a proton is described by two non-interfering sub-processes which differ in neutrino handedness either in the initial or in the final state. 

The first sub-process caused by left-handed neutrinos is standard, i.e. it is due to the exchange of the $Z$-boson. We use the tabulated values of the total cross section $\sigma^{{\rm nc}(Z)}(\omega)$ of this sub-process as function of the energy $\omega$ of the incident neutrinos, which are contained in the work \cite{37}. The rate of the sub-process of the deuteron disintegration due to $Z$-boson exchange $V(nc(Z)|{\rm B})$ is calculated according to a formula similar to (\ref{8}). This rate translates into effective flux of solar neutrinos with their spectrum from the decay of ${}^{8}{\rm B}$, $\Phi_{eff}^{nc(Z)}({}^{8}{\rm B})$, according to the formula
\begin{equation}
V(nc(Z)|{\rm B}) = \Phi_{eff}^{nc(Z)}({}^{8}{\rm B})
\sum_{i=1}^{160}\Delta^{B}p(\omega_{i}^{B})
\sigma^{{\rm nc}(Z)}(\omega_{i}^{B}). 
\label{18}
\end{equation}
We have
\begin{equation}
\Phi_{eff}^{nc(Z)}({}^{8}{\rm B}) = \left\{ 
\begin{array}{ll}
2.16 \cdot 10^{6} \; 
{\rm cm}^{-2}{\rm s}^{-1}, & {\mbox{\rm if}} \hspace{0.3cm} n_{0}=10, \\
2.10 \cdot 10^{6} \; 
{\rm cm}^{-2}{\rm s}^{-1}, & {\mbox{\rm if}} \hspace{0.3cm} n_{0}=11.
\end{array} \right.
\label{19}
\end{equation}

The second sub-process of the deuteron disintegration into a neutron and a proton, caused both left- and right-handed neutrinos, is due to the exchange of the massless pseudoscalar boson. When choosing the method and performing the calculation of the relevant cross section, we take into consideration a number of works on the deuteron disintegration by neutrinos, in particular, the works \cite{37}--\cite{45}. We do not aspire to the degree of accuracy in the choice of the deuteron model and to the precision of calculation which have been achieved in Ref. \cite{37}, \cite{39}--\cite{41}. We are based on the spherically symmetric $np$-potential of the type $U_{0}\delta(r)$ used in Ref. \cite{42}--\cite{45}.

We denote the 4-momenta and their components in the laboratory frame as
\begin{equation}
\nu_{e}(k)+D(P_{0}) \rightarrow \nu_{e}(k')+N_{1}(p_{1})+N_{2}(p_{2}),
\label{20}
\end{equation}
\begin{equation}
k=\{ \omega, {\bf k}\}, \; k'=\{ \omega', {\bf k}'\}, \; P_{0}=\{M_{D}, 
{\bf 0}\}, \; p_{1}=\{ E_{1}, {\bf p}_{1}\}, \; p_{2}=\{ E_{2}, {\bf p}_{2}\}.
\label{21}
\end{equation}
We also introduce the 3-momentum of the center of mass ${\bf P}= {\bf p}_{1}+{\bf p}_{2}$ and the relative 3-momentum of the final nucleons ${\bf p}= ({\bf p}_{1}-{\bf p}_{2})/2$. We describe the deuteron and the final $N_{1}N_{2}$-system state vectors by tensor products (1) of the scalar functions of time and coordinates of the nucleons, and (2) of vectors from the spin space, where the latter are represented by the tensor product of two Dirac bispinors. In turn, the above scalar functions are given by the product (1) of the center-of-mass motion wave functions, and (2) of the functions of the distance $r$ between the nucleons, $\varphi_{D}(r)$ and $\varphi_{N_{1}N_{2}}(r)$. It is accepted that
\begin{equation}
\varphi_{D}(r) = \sqrt{\frac{\gamma}{2\pi}} \cdot \frac{e^{-\gamma r}}{r}, \quad
\varphi_{N_{1}N_{2}}(r) = \frac{\sin(pr+\delta_{0})}{pr}, 
\label{22}
\end{equation}
where $p=|{\bf p}|$, $\gamma = \sqrt{MB}$,  $p\cot\delta_{0} = -1/a_{s}$,  
$(a_{s}^{2}M)^{-1} = 0.0738$ MeV. It is easy to make sure of the following known equality
\begin{equation}
\left| \int d^{3}{\bf r}  \varphi_{D}^{*}(r)\varphi_{N_{1}N_{2}}(r) \right|^{2}
= \frac{8\pi \gamma (1-a_{s}\gamma)^{2}}{(1+a_{s}^{2}p^{2})
(p^{2}+\gamma^{2})^{2}}.
\label{23}
\end{equation}

Let us divide the final state phase space of the deuteron disintegration process into such pairs related to the protons and neutrons, so that the pseudoscalar current with interaction (\ref{1}) be represented in the form
$$J^{5} = \frac{g_{Nps}}{\sqrt{2}} \int d^{3}{\bf r}\left\{ \varphi_{D}^{*}(r)
\left[ \bar{\psi}({\bf p}_{a},M_{p})\gamma^{5}\psi({\bf p}'_{a},M_{p})\cdot
\bar{\psi}({\bf p}_{b},M_{n})\psi({\bf p}'_{b},M_{n}) \right. \right. $$
\begin{equation}
\left. \left. -\bar{\psi}({\bf p}_{a},M_{n})\gamma^{5}\psi({\bf p}'_{a},M_{n})\cdot \bar{\psi}({\bf p}_{b},M_{p})\psi({\bf p}'_{b},M_{p})\right] 
\varphi_{N_{1}N_{2}}(r)\right\}, 
\label{24}
\end{equation}
omitting the mass center motion.
Since the current (\ref{24}) as a function of the mass $M_{n}$ vanishes at $M_{n}=M_{p}$, it is advisable to expand the bilinear forms $\bar{\psi}({\bf p}_{c},M_{n}){\cal O}\psi({\bf p}'_{c},M_{n})$, $c=a,b$, in a Taylor series in powers of $M_{n}-M_{p}$. Restricting to the first power and to the approximation that $E_{i} = M$, we get
\begin{equation}
J^{5} = \frac{g_{Nps}}{\sqrt{2}} \cdot \frac{M_{n}-M_{p}}{M} 
\left[\bar{\psi}({\bf p}_{a},M_{p})\gamma^{5}\psi({\bf p}'_{a},M_{p})\cdot
\bar{\psi}({\bf p}_{b},M_{p})\psi({\bf p}'_{b},M_{p})\right]
\int d^{3}{\bf r} \varphi_{D}^{*}(r)\varphi_{N_{1}N_{2}}(r).
\label{25}
\end{equation}

The factor $1/\sqrt{2}$ in relations (\ref{24}) and (\ref{25}) is necessary to avoid double counting of the cross section due to taking each set of the final states in the phase space determined by the energy-momentum conservation. The phase space considering the kinematics of the deuteron disintegration (\ref{20}) as $2 \rightarrow 3$ process is given by the following formula
\begin{equation}
d^{3}I = \delta(\omega+M_{D}-\omega'-E_{1}-E_{2})\delta({\bf k}-{\bf k}'-
{\bf p}_{1}-{\bf p}_{2}) \; d^{3}{\bf k}' \; d^3{\bf p}_{1} \; d^3{\bf p}_{2}.
\label{26}
\end{equation}
We use the nonrelativistic approximation
\begin{equation}
E_{1}+E_{2}=M_{n}+M_{p}+\frac{{\bf p}_{1}^{2}}{2M}+\frac{{\bf p}_{2}^{2}}{2M}, 
\label{27}
\end{equation}
substitute ${\bf P}$ and ${\bf p}$ for ${\bf p}_{1}$ and ${\bf p}_{2}$ in equation (\ref{26}), and integrate over ${\bf P}$, eliminating the delta-function of 3-momenta. We obtain
\begin{equation}
d^{2}I =  \delta\left( \omega-B-\omega'-\frac{({\bf k}-{\bf k}')^{2}}
{4M}-\frac{p^{2}}{M}\right) \; d^{3}{\bf k}' \; d^3{\bf p}.
\label{28}
\end{equation} 

On the basis of relations (\ref{1}), (\ref{25}) and (\ref{23}), we find the squared modulus of the matrix element $|{\cal M}|^{2}$ for the sub-process of the deuteron disintegration, caused by the massless pseudoscalar boson exchange between the left- and right-handed neutrinos and nucleons. Substituting it into the formula for the differential cross section
\begin{equation}
d\sigma = \frac{|{\cal M}|^{2}}{16\omega \omega' M^{2} (2\pi)^{5}}d^{2}I,
\label{29}
\end{equation} 
using equality (\ref{28}) with knowing the relative energy $E_{r}=p^{2}/M$ and performing a number of integrations, we obtain
$$\sigma^{\rm nc(ps)}(\omega) = \frac{(g_{\nu_{e}ps}g_{Nps})^{2}}
{16\pi^{2}M^{2}}\cdot \left( \frac{M_{n}-M_{p}}{M}\right)^{2}\cdot 
\frac{\sqrt{B} (\sqrt{B}-(a_{s}\sqrt{M})^{-1})^{2}}{\omega}$$
\begin{equation}
\times \int_{0}^{\omega - B} dE_{r}\frac{(\omega - B -E_{r})
\sqrt{E_{r}}}{(E_{r}+B)^{2}(E_{r}+(a_{s}^{2}M)^{-1})}.
\label{30}
\end{equation}

Using the value of the product of the coupling constants of the massless pseudoscalar boson with the electron neutrino and nucleons given below by equality (\ref{33}) and the above values of other constants, we find the cross section values for the discussed deuteron disintegration sub-process for several energies of the incident neutrino $\omega = (2.2+0.2 j)$ MeV, $j=1,2,\ldots,70$. A number of these values is shown in table 5. To ensure that the results of calculations of the cross section $\sigma^{\rm nc(ps)}$ by formula (\ref{30}) are satisfactory, we have carried out calculations of the cross section $\sigma^{\rm nc(Z)}$ in an approach similar to that described above, and we have placed the results of them in the columns "Control" of table 5 along with the accurate enough results, taken from \cite{37}.

\newpage

\begin{center}
{\bf Table 5.} The cross section of the deuteron disintegration \\ sub-processes 
 $\sigma^{\rm nc(Z)}$ и $\sigma^{\rm nc(ps)}$ 
(in units of $10^{-42}$ cm$^{2}$), \\
\begin{tabular}{cccccccc}
\multicolumn{8}{c} {caused by $Z$-boson and massless pseudoscalar boson.} \\ 
\hline
\multicolumn{1}{c}{$\omega$}
&\multicolumn{1}{c}{$\sigma^{\rm nc(Z)}$}  
&\multicolumn{1}{c}{$\sigma^{\rm nc(Z)}$}
&\multicolumn{1}{c}{$\sigma^{\rm nc(ps)}$} 
&\multicolumn{1}{c}{$\omega$}
&\multicolumn{1}{c}{$\sigma^{\rm nc(Z)}$}  
&\multicolumn{1}{c}{$\sigma^{\rm nc(Z)}$}
&\multicolumn{1}{c}{$\sigma^{\rm nc(ps)}$} \\
(MeV) & \cite{37} & Control & Eq. (\ref{30}) 
&(MeV) & \cite{37} & Control & Eq. (\ref{30}) \\
\hline
3.0  & 0.003 & 0.006 & 0.047 & 10.0 & 1.107 & 1.103 & 0.382 \\
4.0  & 0.031 & 0.039 & 0.132 & 11.0 & 1.458 & 1.443 & 0.402 \\
5.0  & 0.096 & 0.108 & 0.201 & 12.0 & 1.860 & 1.831 & 0.418 \\
6.0  & 0.203 & 0.217 & 0.255 & 13.0 & 2.314 & 2.268 & 0.433 \\
7.0  & 0.356 & 0.406 & 0.298 & 14.0 & 2.822 & 2.754 & 0.446 \\
8.0  & 0.557 & 0.568 & 0.331 & 15.0 & 3.382 & 3.290 & 0.457 \\
9.0  & 0.807 & 0.812 & 0.359 & 16.0 & 3.995 & 3.875 & 0.467 \\
\hline
\end{tabular}
\end{center}

We now are able to calculate the rate of the deuteron disintegration into a neutron and a proton, $V(nc(ps)|{\rm B})$, caused by the neutrinos due to the exchange of the massless pseudoscalar boson with nucleons. Doing it by a formula similar to (\ref{8}), where we now put the total flux of left- and right-handed neutrinos $\Phi({}^{8}{\rm B})$ in place of the flux of left-handed neutrinos $0.5 \Phi({}^{8}{\rm B})$. After this, replacing the rate $V(nc(Z)|{\rm B})$ with the rate $V(nc(ps)|{\rm B})$ in the equality (\ref{18}) and keeping for the cross section $\sigma^{{\rm nc}(Z)}$ the status of theoretical cross section calculated on the basis of the standard model of electroweak interactions (as in Ref. \cite{37}), we find the effective solar neutrino flux $\Phi_{eff}^{nc(ps)}({}^{8}{\rm B})$ which is responsible for the rate of the deuteron disintegration  $V(nc(ps)|{\rm B})$. We have
\begin{equation}
\Phi_{eff}^{nc(ps)}({}^{8}{\rm B}) = \left\{
\begin{array}{ll}
(2.90 \pm 0.36) \cdot 10^{6} \; 
{\rm cm}^{-2}{\rm s}^{-1}, & {\mbox{\rm if}} \hspace{0.3cm} n_{0}=10, \\
(2.87 \pm 0.36) \cdot 10^{6} \; 
{\rm cm}^{-2}{\rm s}^{-1}, & {\mbox{\rm if}} \hspace{0.3cm} n_{0}=11,
\end{array}  \right.
\label{31}
\end{equation}
and the uncertainty in (\ref{31}) is only due to the uncertainty in the coupling constant (\ref{33}).

So, the effective flux of solar neutrinos corresponding to the total rate of the two sub-pro- cesses of the neutron disintegration by the neutral currents of the solar neutrinos, $V(nc(Z)|{\rm B})+V(nc(ps)|{\rm B})$, is equal to
\begin{equation}
\Phi_{eff}^{nc}({}^{8}{\rm B}) = \left\{ 
\begin{array}{ll}
(5.06 \pm 0.36) \cdot 10^{6} \; 
{\rm cm}^{-2}{\rm s}^{-1}, & {\mbox{\rm if}} \hspace{0.3cm} n_{0}=10, \\
(4.97 \pm 0.36) \cdot 10^{6} \; 
{\rm cm}^{-2}{\rm s}^{-1}, & {\mbox{\rm if}} \hspace{0.3cm} n_{0}=11,
\end{array} \right.
\label{32}
\end{equation}
that is in good agreement with the above-listed experimental results of SNO.

\begin{center}
{\bf 9. Coupling constants}
\end{center}

Let us find the value of the constant $g_{ps\nu_{e}}g_{psN}$ starting from the point that the electron neutrinos during their movement in the Sun undergo a small number (of order of 10) of collisions with nucleons. Using the density values of the Sun matter varying with the distance from the Sun center as tabulated in Ref. \cite{15}, we find that a tube of 1 cm$^{2}$ cross-section spearing spreading from the center to the periphery of the Sun contains $1.49 \times 10^{12}$ grams of matter and, consequently, $8.91 \times 10^{35}$ nucleons. From here, and on the basis of relation (\ref{4}) and the assumption that a neutrino passes in the Sun through 0.7 to 0.9 (unquestionably, through 0.1 to 1.0) of the amount of matter in the mentioned tube before colliding with a nucleon, we obtain
\begin{equation}
\frac{g_{\nu_{e}ps}g_{Nps}}{4\pi} = (3.2 \pm 0.2)\times 10^{-5}.
\label{33}
\end{equation}
The enough exact value of the discussed coupling constant could be obtained if it were possible to perform extremely complex calculations of the short-term Brownian motion of solar neutrinos in the inhomogeneous, spherically symmetric medium from their production places up to an output for borders of the Sun.

Thus, the product of constants of the postulated interaction of a massless pseudoscalar boson with electron neutrinos and nucleons is by several orders of magnitude smaller than the constants of electromagnetic and weak interactions, respectively $\alpha$ and $g^{2}/4\pi$. Therefore,  the postulated interaction could be named superweak. However, by virtue of that the total cross-section of such a neutrino-nucleon interaction at a low energy, as it is the case for solar neutrinos and reactor antineutrinos, is much larger than that of the standard weak interaction via $Z$-boson exchange, we prefer to call the postulated interaction semi-weak.

\begin{center}
{\bf 10. A few remarks}
\end{center}

At ten-eleven collisions of a solar neutrino with nucleons, the neutrino energy output from the Sun decreases approximately by $0.3\%$ in comparison with the value expected in SSM, what is less than the theoretical uncertainty of the neutrino flux value, $1\%$. This gives us the firm basis to think, that the postulated interaction of stellar neutrinos with nucleons has very little effect on the evolution of this or that star.

The answer to the question about the presence or absence of the interaction of
a massless pseudoscalar boson with a muon neutrino should be sought in experiments with neutral currents of the accelerator neutrinos at small values of the transferred momentum square.

\begin{center}
{\bf 11. Conclusion}
\end{center}

It is justified to think that the range of phenomena potentially involved in the manifestation of the postulated new interaction will expand with time. In our opinion, it is first of all necessary to pay close attention to setting up  experiments with reactor antineutrinos. We expect that, in an experiment on the deuteron disintegration by the neutral currents, the contribution of the semi-weak interaction to the observed rate of events will be about three times greater than the contribution of the electroweak interaction \cite{46}. A bright manifestation of the new interaction is expected in observing the splitting rate of a number of light stable nuclei (He-3, Li-7, Be-9, and F-19) by reactor antineutrinos, where the contribution from electroweak interaction may be exceeded by approximately six orders of magnitude \cite{47}. It also seems appropriate to clarify the consequences of interaction (\ref{1}) in various astrophysical processes.

I am sincerely grateful to S.P. Baranov, A.M. Snigirev and I.P. Volobuev for useful discussions on some of the problems anyhow concerning the present work.

\end{small}
\end{document}